\documentclass[10pt]{article}
\usepackage{amssymb}
\usepackage{amsthm}
\usepackage{authblk}
\usepackage{setspace}
\usepackage{hyperref}
\usepackage[textwidth=6.5in,textheight=8.25in,centering]{geometry}
\usepackage[pdftex]{graphicx}

\onehalfspace
\begin{document}
\begin{center}
{\LARGE{\textbf{Generalization of Phonon Confinement Model for Interpretation of Raman Line-Shape from Nano-Silicon}}}

\vspace{0.5 cm}
\textit{Manushree Tanwar $ ^{1} $, Priyanka Yogi $^{1}$, Simran Lambora $ ^{1}$ Suryakant Mishra $ ^{1}$, Shailendra K. Saxena $ ^{1,2}$,  , Pankaj R. Sagdeo $ ^{1}$, Alexander S. Krylov $ ^{3}$ and Rajesh Kumar}\footnote{Corresponding author email: rajeshkumar@iiti.ac.in; \href{http://magse.webs.com/} {\textbf{http://magse.webs.com/}}}

\vspace{0.5 cm}
$ ^{1}$ Material Research Laboratory, Discipline of Physics \& MEMS, Indian Institute of Technology Indore, Simrol-453552, India

$ ^{2}$ national Institute for nanotechnology, University of Alberta, Edmonton, Canada

$ ^{3}$ Kirensky Institute of Physics, Federal Research Center KSC SB RAS

\vspace{1 cm}
ABSTRACT

\end{center}
 A comparative analysis of two Raman line-shape functions has been carried out to validate the true representation of experimentally observed Raman scattering data for semiconducting nanomaterials.  A modified form of already existing phonon confinement model incorporates two basic considerations, phonon momentum conservation and shift in zone centre phonon frequency.   After incorporation of the above mentioned two factors, a rather symmetric Raman line-shape is generated which is in contrary to the usual asymmetric Raman line-shapes obtained from nanostructured semiconductor. By fitting an experimentally observed Raman scattering data from silicon nanostructures, prepared by metal induced etching, it can be established that the Raman line-shape obtained within the framework of phonon confinement model is a true representative Raman line-shape of sufficiently low dimensions semiconductors. 
\vspace{0.5cm}


\section{Introduction}

Raman scattering [1–-4], discovered in 1928, remains one of the versatile methods  for study of materials in its all states including crystalline and nano-crystalline forms.  Apart from knowing crystal structure, chemical composition, crystal defects, it also investigates different physical phenomena taking place at microscopic levels like confinement and Fano resonance etc.[4–-9] Raman scattering proves to be of superior sensitivity over other characterization tools especially where a quick and extremely sensitive technique is required especially in low dimensional semiconductors.  It is an established fact that for solid semiconductor materials, Raman scattering occurs from only the zone center phonons[4,10,11] which results in a symmetric Raman spectrum with peak position corresponding to zone center phonon. In the nanomaterials, this rule is relaxed enabling phonons other than zone center also to participate in the Raman scattering. As a result, a red-shifted Raman spectrum with asymmetric broadening towards lower wavenumber side are observed in low dimensional materials like nanowires or in general nanostructures (NSs)[12] where phonons and electrons are confinement. Richter et al[13] has given phonon confinement model (PCM) which was modified by Campbell et al[14] later to incorporate some observations into the form of a line-shape function. This model enables one to estimate NSs size. Phonon confinement effect in terms of asymmetric Raman line shape has been observed earlier by various groups[15–-17] working in this field.

The above mentioned PCM allows one to incorporate various other effects, affecting the Raman scattering, by suitably modifying the Raman line-shape Eq.[12,18–-20]. As described earlier, the origin of the Raman line-shape equation is the relaxation in the zone-center phonon selection rule and confinement induced uncertainty in the wave vector of phonons. This is done by appropriately choosing weighing functions and integration limits on the wave-vector. Due to unavailability of information regarding the exact shape, size and size-distribution it is not always easy to use an unambiguous weighing function thus result in discrepancy between theoretical Raman line-shape and experimentally observed Raman scattering data. Besides certain disadvantages, the PCM has been used as the most widely used formulation when it comes to interpretation of the Raman scattering results obtained from low dimensional semiconductors especially elemental ones. In recent attempts to rectify the shortcomings of the PCM, Jia et al[21] modified the equation predicted by PCM and propose as new Raman line-shape equation in the same regime of low dimensions. Main aim of this paper is to compare the modified model (MPCM) with the original PCM to carry out a fare assessment on the success of the MPCM in explaining the experimental Raman scattering results. 
 
\section{Experimental Details}

Commercially available boron doped p type Si wafer of resistivity 0.01 $\Omega$-cm has been used to fabricate SiNSs using metal induced etching (MIE). Cleaned Si wafers are immersed in 4.8 M HF solutions containing 5mM AgNO$_3$ for Ag deposition. The Ag deposited Si wafer then are transferred in the etching solution which contains 4.8M HF \& 0.5M H$_2$O$_2$. In order to remove the metal nano particles the etched Si wafer is immersed in the HNO$_3$ solution. During the HNO$_3$ treatment strong oxide layer is formed on SiNSs, to remove this oxide layer induced by HNO$_3$ the samples are dipped again in HF solution. The details of MIE mechanism have been reported in the previous literature [22–-24]. In order to confirms the wire like structures formed after etching, transmission electron microscopy (TEM) was carried out using TEM- Gatan model 636MA.  Raman spectra were recorded using a large focal length micro-Raman spectrometer (Horiba JobinYvon) in back scattering geometry with 2.54 eV excitation laser with a beam waist of $\sim$ 1$\mu$m and charge coupled device (CCD) detector with resolution of $\sim$ 0.5 cm$^{-1}$.

\section{Results and Discussion}
Figure 1 shows theoretical Raman line shapes generated using Eq.1, the modified phonon confinement model[21] (MPCM), for different crystallite size of Si for one dimensional confinement. It is evident from Figure 1 that as the size of NS decreases the Raman peak shifts towards the lower wavenumber (red-shift). Inset of Figure 1 shows the asymmetry ratio (defined as $\alpha_\rho = \frac{\gamma_l}{\gamma_l}$ where $\gamma_l$ and $\gamma_l$ are the lower and higher spectral half widths respectively of the Raman line-shape) and red-shift in peak position of Raman line shape as a function of NS size. The size dependent variation in asymmetry ratio and peak position has been shown in red and green symbols respectively in the inset (Figure 1).
\begin{equation}
I(\omega , D) \propto \rho(\omega) \frac{1}{\pi D^3} \int _{2\pi -1} ^{2\pi +1} \frac{4 \pi Q^2\left[3\frac{sin \left(\frac{Q}{2}\right)}{\pi ^3 Q (4\pi ^2 - Q^2)}  \right]^2 \left( \frac{\Gamma}{2}\right)}{[ \omega - \omega^{'}(Q) ]^2+\left(\frac{\Gamma}{2} \right) ^2} dQ
\end{equation}

where, $\omega ' (Q) = 521 \left( 1-0.23 \left(\frac{Qa}{2 \pi D}\right)^2 \right)$ is the dispersion relation and $\rho (\omega) \sim \frac{n(\omega)+1}{\omega} = \frac{\left( e^{\frac{\hbar \omega}{kT}}-1\right)^{-1}+1}{\omega}$ is the Bose-Einstein occupation number, $\Gamma$ is the line width Raman line shape and `D' is the crystallite size in nm.

Three key observations, which are used to identify presence of confinement effect in Si, are first, red-shifted Raman line-shape with respect to the crystalline silicon (c-Si), second, the non-unity asymmetry ratio and third, the broadened Raman line shape as compared to its c-Si counterpart. It is evident from Figure 1 that as the size of SiNS decreases from 10 nm to 2 nm the Raman line shape exhibits red shift with respect to the c-Si. However, this observation is a must for systems with confinement effect but may also arise from other effects like stress and temperature[25–29]. The broadening of the Raman line-shape is also increasing with decreasing size but no apparent change in asymmetry is observed with decreasing size (Figure 1, inset). The symmetry of the Raman line-shape remains intact even for NS size of 2 nm which is far below the Bohr’s exciton radius of Si. This is very unusual Raman spectra from Si NSs of sizes less than 10 nm. Thus Eq. 1, though has merits, does not seem to be the representative of a typical Raman spectral line-shape for low dimensional Si. A discrepancy between the Raman line-shape represented using Eq. 1 and experimentally observed Raman scattering data is also observed and will be discussed later on.

\begin{figure}
\begin{center}
\includegraphics[width=9cm]{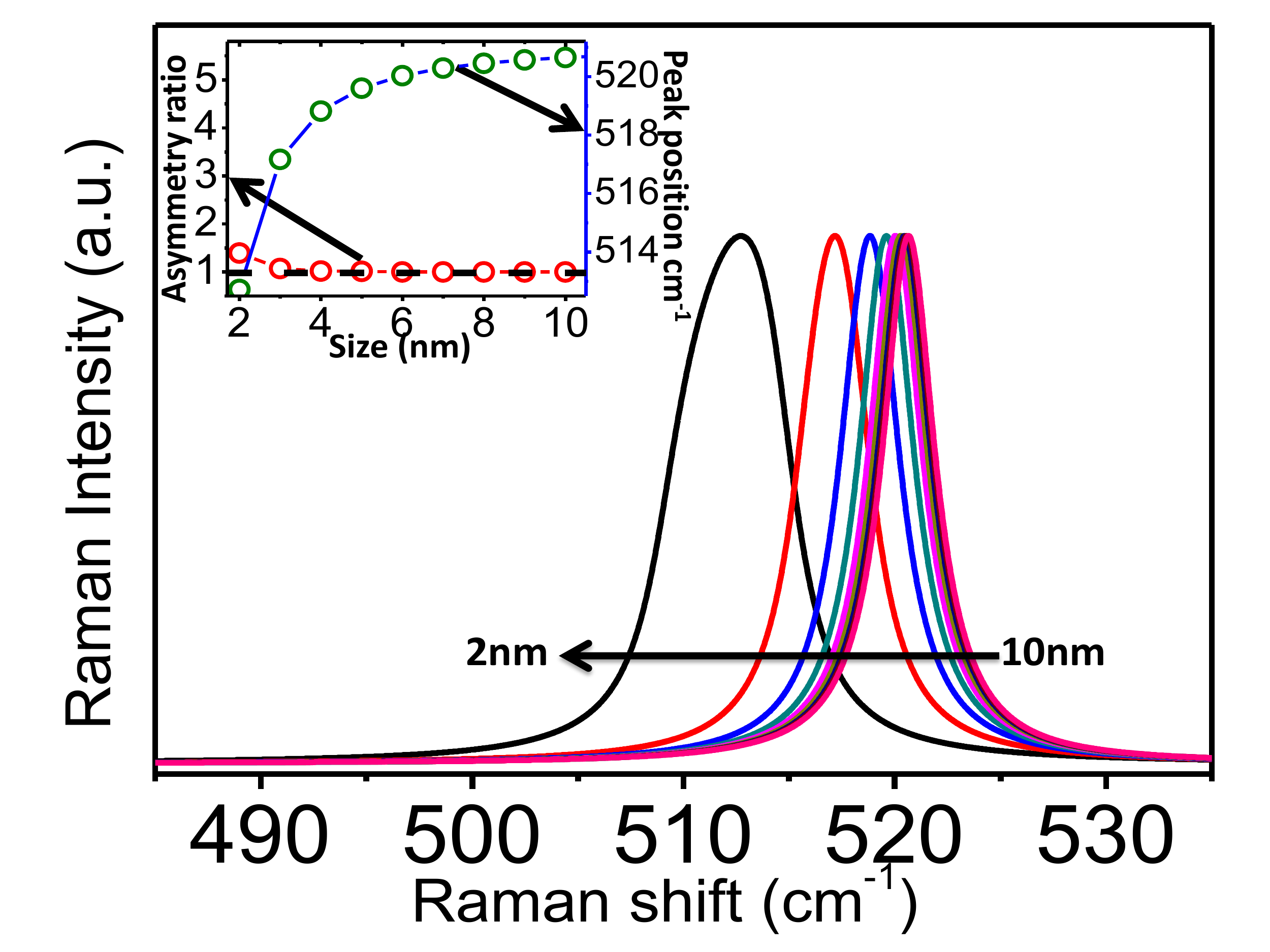}
\caption{Raman line-shapes obtained using modified phonon confinement model (Eq. 1) for different nano crystallite sizes.}
\end{center}
\end{figure}

For comparison, Raman line-shape obtained using well-established PCM[13,16,17,23,30], as represented by Eq 2 below, is also analysed and compared with the one obtained using the modified model (Eq. 1). Figure 2 shows the Raman line shapes generated by PCM using Eq. 2 for different nano structure size. It is evident from Figure 2 that as the size of nano structure decreases from 10 nm to 2 nm the broadening and asymmetry of Raman line shape increases along with spectral red-shift. It is also evident from the Raman line shape generated using PCM that the peak position is more red shifted for 2nm size as compared to the 10nm size.To further confirm the asymmetric line shape we calculate the asymmetry ratio using Figure 1 and Figure 2. Comparison between the insets of Figure 1 and Figure 2 clearly shows how Eq. 2 (PCM) results in asymmetric Raman line-shapes with increasing asymmetry ratio with size. This is a true nature of Raman scattering from low dimensional Si. In contrary, Eq. 1 (MPCM) results in  a rather symmetric Raman line-shape a non-representation of experimentally observed Raman scattering from Si NSs.

\begin{equation}
I_1(\omega , D) \propto \int _0 ^1 \frac{e^{-\frac{k^2 D^2}{4a^2}}}{[\omega -\omega (k)]^2+ \left( \frac{\gamma}{2} \right)^2} d^n k
\end{equation}
 where, $I_1(\omega , D)$ shows the Eq. for PCM Raman line-shape with $\omega = \sqrt{171400+100000 cos \left(\frac{\pi k}{2}\right)}$ is the phonon dispersion relation of Si, $k$ is reduced wave vector, `$a$’ being lattice parameter of material, `D’ denotes the size of NS present in the sample, $\Gamma$ is the FWHM of Raman spectrum of the bulk material and `$n$’ being the order of confinement.

\begin{figure}
\begin{center}
\includegraphics[width=9cm]{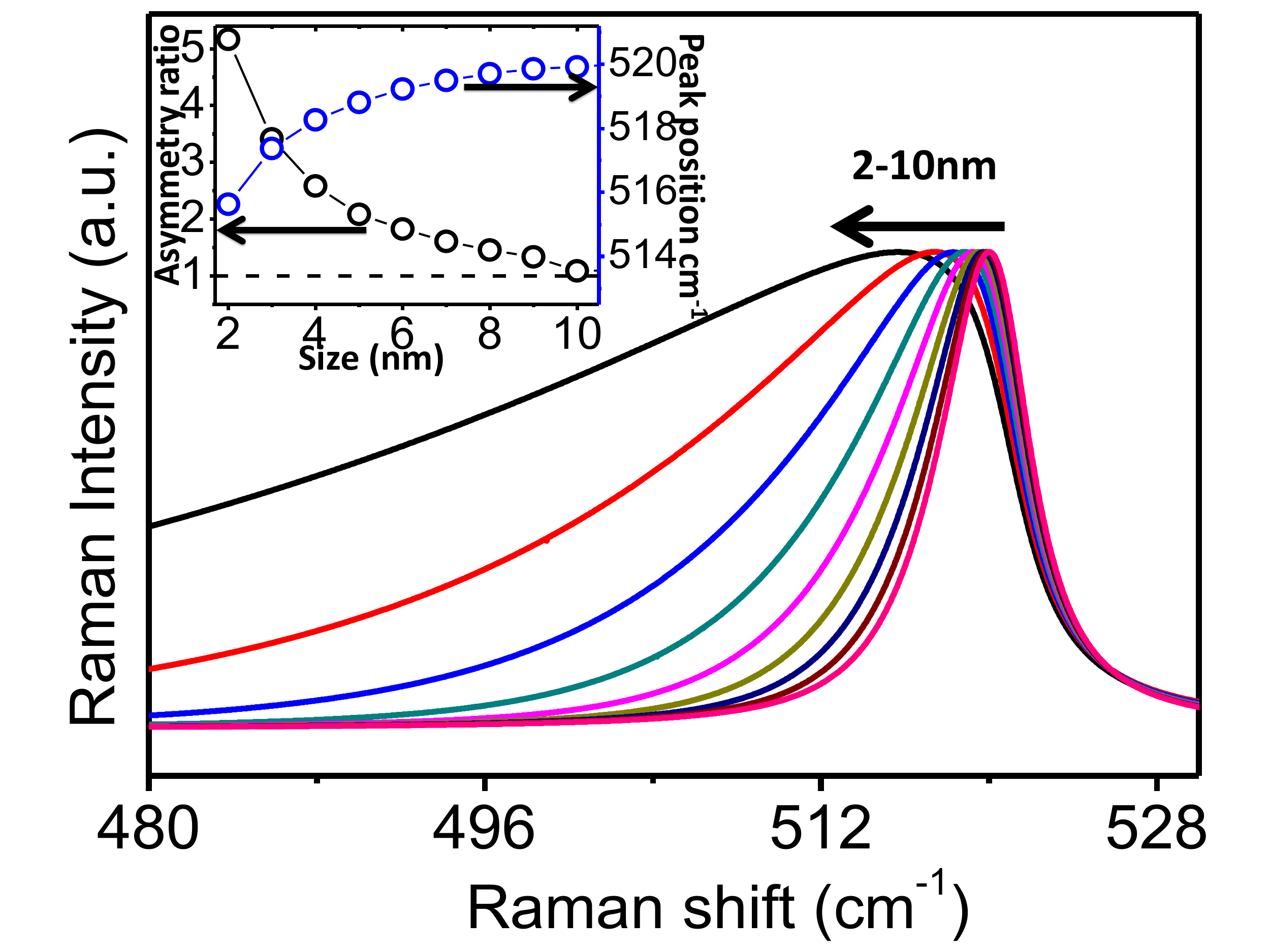}
\caption{Raman line-shapes theoretically generated using phonon confinement model Eq. 2 for different SiNSs.}
\end{center}
\end{figure}

Discrete points in Figure 3 (blue as well as red) shows the experimentally obtained data whereas the solid lines in red and blue are the line-shapes generated using Eq. 1 and Eq. 2 respectively. As can be seen clearly from Figure 3, the experimental Raman data does not fit with the theoretically obtained line-shape when crystallite size of 7 nm is used in Eq. 1 (MPCM). In fact, we have also try to fit the Raman line-shape obtained using different sizes between 2nm-10 nm which also do not show fit with the experimental data. On the other hand, the experimental Raman scattering data shows a good agreement with the line-shape generated using Eq 2 (PCM) for a crystallite size of 7 nm as shown in blue curve (Figure 3). The TEM image of one of the Si nanowires present in the sample is shown in the inset of Figure 3. A size of approximately 40 nm is seen in the TEM image which is not in agreement with the size estimated using Raman scattering. This discrepancy appears due to the fact that the nanocrystals responsible for giving rise to asymmetric Raman line-shape are present inside the porous structures of the nanowires and are not resolved using electron microscopes. The smaller NSs size obtained using Raman data are the actual sizes as the Raman scattering is more sensitive in this regard as compared to other techniques[5,31]. More discussion about this discrepancy has been reported somewhere else [33].
 
\begin{figure}
\begin{center}
\includegraphics[width=12cm]{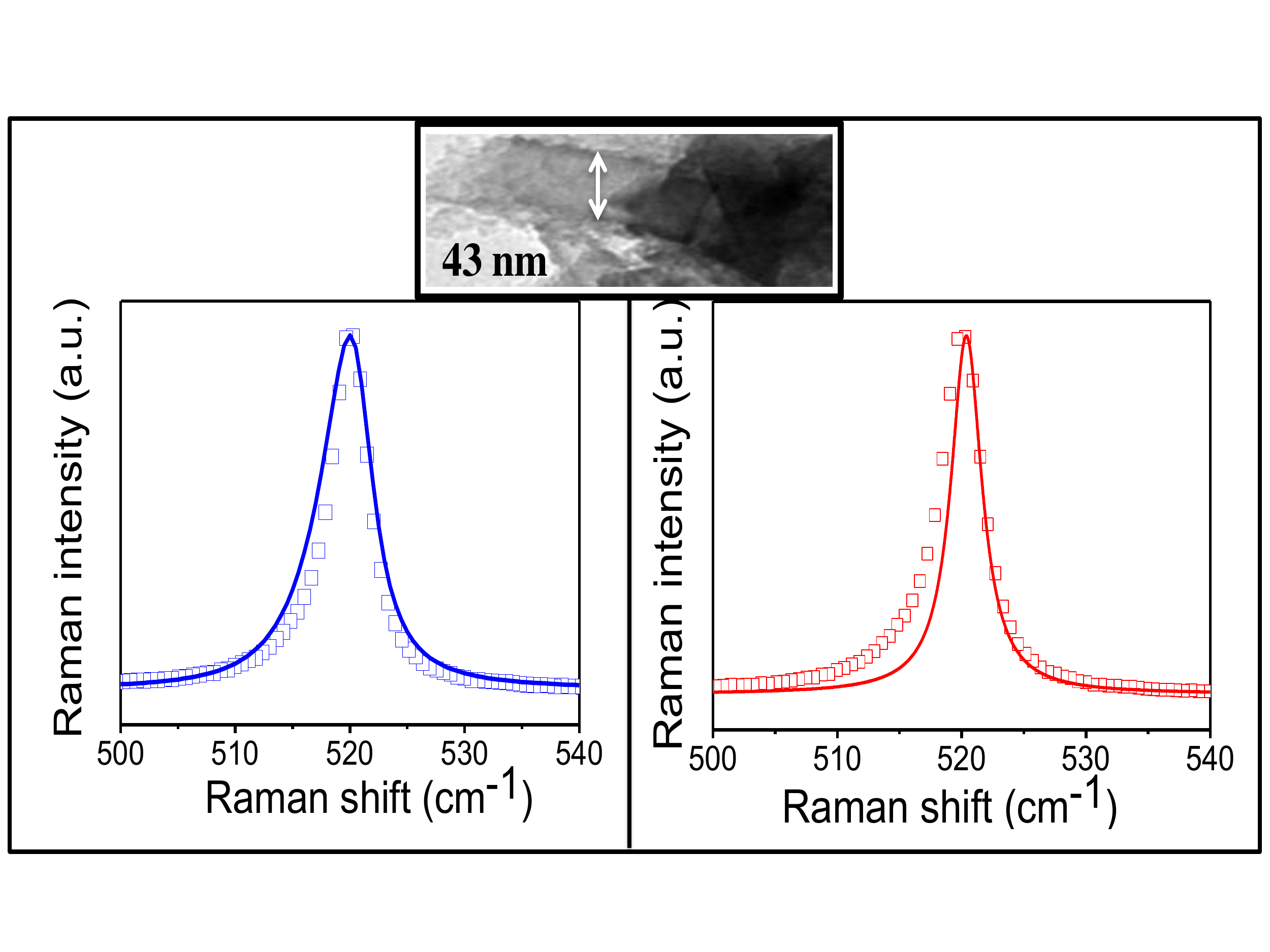}
\caption{Fitting of experimentally observed Raman scattering data (discreet points) with Eq. 1 (red curve) and Eq.2 (blue curve). Inset shows the TEM image of silicon nanowires present in the sample.}
\end{center}
\end{figure}  

It is very clear from above discussion that the MPCM does not represent the actual Raman scattering phenomenon taking place from low dimensional Si whereas PCM still appears to be the more generalized form for representation of the Raman line-shapes. The reasons for the incompleteness of Eq. 1 include the non-consideration of the participation of absolute zone-centre phonons in the Raman scattering from nano-Si. Rather MPCM considers that the zone centre of the phonon dispersion gets shifted due to finite size similar to the shifting of the band edge as discussed in the context of size dependent band gap enhancement in nano-Si. It is worth mentioning here that, considering the shift of the phonon dispersion zone centre all-together sounds a nice approach for explanation of Raman line-shapes at nanoscale but at the same time appears incomplete. Another possible advantage of using Eq. 1 is the consideration of partial phonon dispersion curve to take care of the momentum conservation. This also, though conceptually correct, leads to the discrepancy between the Eq. 1 and experimental Raman scattering data. A better way to consider the zone centre shift and momentum conservation needs to be explored but in the absence of it, PCM still seems the best and more generalized form of the Raman line-shape from nano-Si. 

\section{Conclusions}
Comparison of Raman line-shapes generated using phonon confinement model (PCM) with a modified version (MPCM) of the same reveals certain discrepancies between the two for the case of SiNSs. The MPCM has been generated by incorporating the shift in the zone centre phonon at low dimensions and phonon momentum conservation which is the advantage of the model. The MPCM remains symmetric even for extremely low NSs’ size which is not a true representation of experimentally observed Raman scattering data. On the other hand the unmodified PCM still fits well with the experimental data and appears to be the more generalized form of Raman line-shape in low dimensional regime though attempts may be taken for more generalization of the PCM appropriately.

\subsection*{Acknowledgement} 
The authors thank Dr. N. P. Lalla (UGC−-DAE consortium for Scientific Research, Indore) for TEM measurements and Dr. V. G. Sathe (UGC−-DAE consortium for Scientific Research, Indore) for Raman measurements. Thanks to Dr. J. Jayabalan (RRCAT, Indore) and Ms. Rupsa (IIT Indore) for useful discussion. Financial support from the Department of Science and Technology (DST), Government of India, is also acknowledged. The authors thank MHRD for providing a fellowship.

\newpage


\begin{thebibliography}{33}

\bibitem{1}	C. Raman, A new radiation, Indian J. Phys. 02 (1928), .
\bibitem{2}	C.V. Raman and K.S. Krishnan, A New Type of Secondary Radiation, Nature 121 (1928), pp. 501–502.
\bibitem{3}	D. Abidi, B. Jusserand and J.-L. Fave, Raman scattering studies of heavily doped microcrystalline porous silicon and porous silicon free-standing membranes, Phys. Rev. B 82 (2010), pp. 075210.
\bibitem{4}	G. Gouadec and P. Colomban, Raman Spectroscopy of nanomaterials: How spectra relate to disorder, particle size and mechanical properties, Prog. Cryst. Growth Charact. Mater. 53 (2007), pp. 1–56.
\bibitem{5}	C.M. Hessel, J. Wei, D. Reid, H. Fujii, M.C. Downer and B.A. Korgel, Raman Spectroscopy of Oxide-Embedded and Ligand-Stabilized Silicon Nanocrystals, J. Phys. Chem. Lett. 3 (2012), pp. 1089–1093.
\bibitem{6}	P. Yogi, S. Mishra, S.K. Saxena, V. Kumar and R. Kumar, Fano Scattering: Manifestation of Acoustic Phonons at the Nanoscale, J. Phys. Chem. Lett. 7 (2016), pp. 5291–5296.
\bibitem{7}	D.M. Sagar, J.M. Atkin, P.K.B. Palomaki, N.R. Neale, J.L. Blackburn, J.C. Johnson et al., Quantum confined electron-phonon interaction in silicon nanocrystals, Nano Lett. 15 (2015), pp. 1511–1516.
\bibitem{8}	G. Bepete, A. Pénicaud, C. Drummond and E. Anglaret, Raman Signatures of Single Layer Graphene Dispersed in Degassed Water, ““Eau de Graphene”’,” J. Phys. Chem. C 120 (2016), pp. 28204–28214.
\bibitem{9}	K. Jin, S. Pan and G. Yang, Fano effect of resonant Raman scattering in a semiconductor quantum well, Phys. Rev. B 50 (1994), pp. 8584–8588.
\bibitem{10}		F. Cerdeira, T.A. Fjeldly and M. Cardona, Effect of Free Carriers on Zone-Center Vibrational Modes in Heavily Doped p-type Si. II. Optical Modes, Phys. Rev. B 8 (1973), pp. 4734–4745.
\bibitem{11}		T.A. Fjeldly, F. Cerdeira and M. Cardona, Effects of Free Carriers on Zone-Center Vibrational Modes in Heavily Doped $p$-type Si. I. Acoustical Modes, Phys. Rev. B 8 (1973), pp. 4723–4733.
\bibitem{12}		R. Kumar, G. Sahu, S.K. Saxena, H.M. Rai and P.R. Sagdeo, Qualitative Evolution of Asymmetric Raman Line-Shape for NanoStructures, Silicon 6 (2014), pp. 117–121.
\bibitem{13}		H. Richter, Z.P. Wang and L. Ley, The one phonon Raman spectrum in microcrystalline silicon, Solid State Commun. 39 (1981), pp. 625–629.
\bibitem{14}		I.H. Campbell and P.M. Fauchet, The effects of microcrystal size and shape on the one phonon Raman spectra of crystalline semiconductors, Solid State Commun. 58 (1986), pp. 739–741.
\bibitem{15}		S. Piscanec, M. Cantoro, A.C. Ferrari, J.A. Zapien, Y. Lifshitz, S.T. Lee et al., Raman spectroscopy of silicon nanowires, Phys. Rev. B 68 (2003), pp. 241312.
\bibitem{16}		K.W. Adu, H.R. Gutiérrez, U.J. Kim, G.U. Sumanasekera and P.C. Eklund, Confined Phonons in Si Nanowires, Nano Lett. 5 (2005), pp. 409–414.
\bibitem{17}		N. Fukata, T. Oshima, K. Murakami, T. Kizuka, T. Tsurui and S. Ito, Phonon confinement effect of silicon nanowires synthesized by laser ablation, Appl. Phys. Lett. 86 (2005), pp. 213112.
\bibitem{18}		I.H. Campbell and P.M. Fauchet, The effects of microcrystal size and shape on the one phonon Raman spectra of crystalline semiconductors, Solid State Commun. 58 (1986), pp. 739–741.
\bibitem{19}		F.J. Bartoli and T.A. Litovitz, Analysis of Orientational Broadening of Raman Line Shapes, J. Chem. Phys. 56 (1972), pp. 404–412.
\bibitem{20}		P. Yogi, S.K. Saxena, S. Mishra, H.M. Rai, R. Late, V. Kumar et al., Interplay between phonon confinement and Fano effect on Raman line shape for semiconductor nanostructures: Analytical study, Solid State Commun. 230 (2016), pp. 25–29.
\bibitem{21}		X. Jia, Z. Lin, T. Zhang, B. Puthen-Veettil, T. Yang, K. Nomoto et al., Accurate analysis of the size distribution and crystallinity of boron doped Si nanocrystals via Raman and PL spectra, RSC Adv. 7 (2017), pp. 34244–34250.
\bibitem{22}		S.K. Saxena, P. Yogi, P. Yadav, S. Mishra, H. Pandey, H.M. Rai et al., Role of metal nanoparticles on porosification of silicon by metal induced etching (MIE), Superlattices Microstruct. 94 (2016), pp. 101–107.
\bibitem{23}		P. Yogi, D. Poonia, S. Mishra, S.K. Saxena, S. Roy, V. Kumar et al., Spectral Anomaly in Raman Scattering from p-Type Silicon Nanowires, J. Phys. Chem. C 121 (2017), pp. 5372–5378.
\bibitem{24}		K. Peng, A. Lu, R. Zhang and S.-T. Lee, Motility of Metal Nanoparticles in Silicon and Induced Anisotropic Silicon Etching, Adv. Funct. Mater. 18 (2008), pp. 3026–3035.
\bibitem{25}		S. Ves, I. Loa, K. Syassen, F. Widulle and M. Cardona, Raman Lineshapes of GaP under Pressure, Phys. Status Solidi B 223 (2001), pp. 241–245.
\bibitem{26}		N.H. Nickel, P. Lengsfeld and I. Sieber, Raman spectroscopy of heavily doped polycrystalline silicon thin films, Phys. Rev. B 61 (2000), pp. 15558–15561.
\bibitem{27}		J. Anaya, A. Torres, V. Hortelano, J. Jiménez, A.C. Prieto, A. Rodríguez et al., Raman spectrum of Si nanowires: temperature and phonon confinement effects, Appl. Phys. A 114 (2013), pp. 1321–1331.
\bibitem{28}		K.W. Adu, H.R. Gutiérrez, U.J. Kim and P.C. Eklund, Inhomogeneous laser heating and phonon confinement in silicon nanowires: A micro-Raman scattering study, Phys. Rev. B 73 (2006), .
\bibitem{29}		S. Bhattacharyya, D. Churochkin and R.M. Erasmus, Anomalous Raman features of silicon nanowires under high pressure, Appl. Phys. Lett. 97 (2010), pp. 141912–141912–3.
\bibitem{30}		I.H. Campbell and P.M. Fauchet, The effects of microcrystal size and shape on the one phonon Raman spectra of crystalline semiconductors, Solid State Commun. 58 (1986), pp. 739–741.
\bibitem{31}		C.L. Baldwin, N.W. Bigelow and D.J. Masiello, Thermal Signatures of Plasmonic Fano Interferences: Toward the Achievement of Nanolocalized Temperature Manipulation, J. Phys. Chem. Lett. 5 (2014), pp. 1347–1354.
\bibitem{32}		K.W. Adu, Q. Xiong, H.R. Gutierrez, G. Chen and P.C. Eklund, Raman scattering as a probe of phonon confinement and surface optical modes in semiconducting nanowires, Appl. Phys. A 85 (2006), pp. 287–297.
\bibitem{33}	S.K. Saxena, P. Yogi, S. Mishra, H.M. Rai, V. Mishra, M.K. Warshi et al., Amplification or cancellation of Fano resonance and quantum confinement induced asymmetries in Raman line-shapes, Phys. Chem. Chem. Phys. (2017), .




\end{thebibliography}
\end{document}